\documentclass[twocolumn,showpacs,preprintnumbers,amsmath,amssymb,fleqn]{revtex4} 
\usepackage{amsmath,amssymb,graphicx}
\begin{document}

\title{Nonreciprocal magnon blockade based on nonlinear effects}
\author{Han-Qiu Zhang$^{1}$}
\author{Shuang-Shuo Chu$^{1}$}
\author{Jian-Song Zhang$^{1}$}
\email{jszhang1981@zju.edu.cn}
\author{Wen-Xue Zhong$^{1}$}
\author{Guang-Ling Cheng$^{1}$}
\email{glingcheng@ecjtu.edu.cn}
\affiliation{$^{1}$Department of Applied Physics, East China Jiaotong University,
Nanchang 330013, People's Republic of China}

\begin{abstract}
We present an alternative scheme to achieve nonreciprocal unconventional magnon blockade (NUMB) in a hybrid system formed by two microwave cavities and one yttrium iron garnet (YIG) sphere, where the pump and signal cavities interact nonlinearly with each other and the signal cavity is coupled to the YIG sphere. It is found that the nonlinear coupling occurs between the pump cavity and magnon modes due to the dispersive interactions among three bosonic modes. Meanwhile, the Kerr nonlinearity is present in the pump cavity. Based on these nonlinear effects, a nonreciprocal magnon blockade could be achieved with the help of weak parametric driving of the pump cavity. The present work provides an alternative method to prepare single magnon resource, which may be helpful for quantum information processing.
\end{abstract}
\maketitle

Significant endeavors have been devoted to the study of photon blockade (PB) \cite{Imamoglu1997,Miranowicz2013}. The PB, one of the schemes to generate strong antibunching photons, which is a phenomenon when a first single photon in a cavity blocks the transmission of a second one \cite{Miranowicz2013}. This quantum phenomenon creates single-photon sources which are useful for quantum information and communication technology \cite{Birnbaum2005}. The PB has been theoretically predicted in a variety of systems, such as circuit cavity quantum electrodynamics systems \cite{Hoffman2011} and photonic crystal systems with quantum dots \cite{Faraon2008}. Quantum destruction interference induced unconventional photon blockade (UPB) can provide a strong PB effect even when the coupling strength is weak \cite{Shen2015,Liew2010,Flayac2013,Flayac2017,Bamba2011}. To date, various systems are proposed to accomplish the UPB, such as quantum well systems \cite{Kyriienko2014}, two-tunnel coupled cavity systems \cite{Zhou2015,Xu2014}, as well as optomechanical systems \cite{Sarma2018}.
\par On the other hand, hybrid magnetic subsystems based on collective excitation of spin waves in ferromagnetic materials provide powerful platforms for realizing various quantum effects \cite{Wang2023}. Recently, the YIG sphere, which has attracted considerable attention due to its unique properties such as extremely high spin density, low damping, high spin density, and flexible tunability \cite{Bourhill2016}. Therefore, the hybrid magnonical systems provide a promising platform to implement quantum information tasks \cite{Liu2019,Wu2021,Li2018,Zheng2023,Chen2023}. The UMB can be realized in various systems, such as qubit-magnon hybrid quantum systems \cite{Xie2020,Fan2023} and magnon-atom hybrid quantum systems \cite{WangF2022,Yan2024}. Nonreciprocal devices are crucial in information transmission \cite{Bino2018,Xie2022}. Very recently, the authors of Ref. \cite{Huang2024} proposed a scheme to realize the NMB based on the Barnett effct in a magnon-based hybrid system. They found that the nonreciprocity, which is indispensable for achieving quantum nonreciprocal magnetic devices, can be obtained by controlling the direction of magnetic field.\setlength{\parskip}{0pt}
\par In this Letter, we study how to implement the NUMB in a hybrid nonlinear cavity-magnon system including a pump cavity, a signal cavity and magnon in YIG sphere. Under the larger detuning conditions, the nonlinear interaction between pump cavity and magnon could be achieved due to the transfer of photon-photon nolinearity through magnetic dipole interaction between the signal cavity and magnon. Meanwhile the Kerr effect of pump cavity could be also obtained and the signal of Kerr parameter can be adjusted via changing the detuning of the driving field from the signal cavity. Different from the general approach of achieving nonreciprocity via the Sagnac-Fizeau effect \cite{Xie2022} or the Barnett effect \cite{Huang2024}, the nonreciprocity is achieved by changing the effective Kerr nonlinearity from positive to negative values. We find two optimal conditions for minimizing the second-order correlation function with the help of the analytical expression. The NUMB with weak coupling and driving may have potentially significant applications in quantum optics and hybrid quantum networks.
\begin{figure}[tbp]
\centering {\scalebox{0.2}[0.2]{\includegraphics{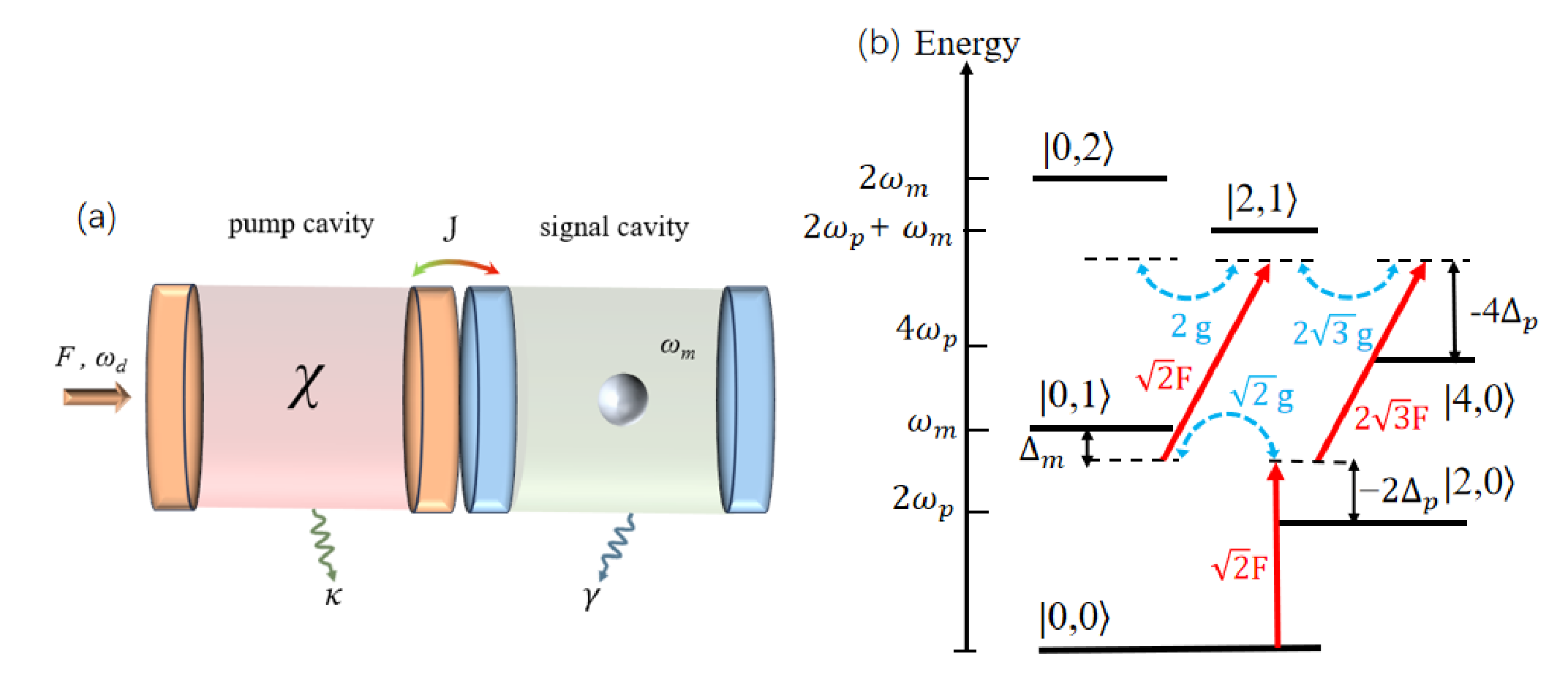}}}
\caption{(a) Sketch of a hybrid cavity magnonic system, where the pump cavity is driven coherently by a field $F$, and a YIG sphere with frequency $\omega_{m}$.
(b) Energy-level diagram. The corresponding transition pathways show the destructive interference process.}\label{fig1}
\end{figure}
\par As schematically shown in Fig. 1, we consider a hybrid cavity magnonic system consisting of two microwave cavities and one YIG sphere. One is a pump cavity with frequency $\omega_{p}$, and the other is a signal cavity with frequency $\omega_{s}$, nonlinearly coupled through second-order nonlinearity $\chi^{\left(  2\right)  }$ \cite{Fang2023}. The magnon mode in the YIG sphere couples with the signal cavity through magnetic dipole interaction. The pump cavity is driven coherently which is essential for the present scheme. External driving for the pump cavity is essential for the scheme to work. Therefore, the Hamiltonian of such a system is given by
\begin{equation}H=H_0+V,\end{equation}
\begin{equation} H_0=\omega_{p}a_{p}^{\dagger}a_{p}+\omega_{s}a_{s}^{\dagger}a_{s}+\omega_{m}m^{\dagger}m,\end{equation}
\begin{equation}\begin{aligned}
V&=g_{ms}\big(a_{s}m^{\dagger}+a_{s}^{\dagger}m\big)+J\Big(a_{p}^2a_{s}^{\dagger}+a_{p}^{\dagger2}a_{s}\Big)\\ &\quad+F\Big(a_{p}^2e^{i\omega_{d}t}+a_{p}^{\dagger2}e^{-i\omega_{d}t}\Big),
\end{aligned}\end{equation}
where $H_0$ is the free Hamiltonian of the pump cavity.
Here $\omega_{p}$,  $\omega_{s}$, and $\omega_{m}$ are the frequencies of the pump cavity, signal cavity, and magnon mode, $a_k(a_k^{\dagger})(k=p,s)$ represents the photon annihilation (creation) operator of the pump and signal cavities, and $m(m^{\dagger})$ is the annihilation (creation) operator of magnon mode. $F$ and $\omega_{d}$ are the amplitude and frequency of the driving field. Then $J$ is the intercavity parametric coupling coefficient, $g_{ms}$ denotes the coupling rate between the magnon mode and the signal mode. For convenience, we study the dynamics of such a system in a rotating reference frame defined by the unitary operator $U=\exp\left[\frac{i\omega_{d}t}{2}(a_p^\dagger a_p+2a_s^\dagger a_s+2m^\dagger m)\right]$, the Hamiltonian of the system at the driving frequency $\omega_{d}$ can be written as
\begin{equation}\begin{aligned}H_R&=\Delta_pa_p^\dagger a_p+\Delta_sa_s^\dagger a_s+\widetilde{\Delta}_mm^\dagger m+J\big(a_p^2a_s^\dagger+a_p^{\dagger2}a_s\big)\\&\quad+g_{ms}\big(a_sm^\dagger+a_s^\dagger m\big)+F\big(a_p^2+a_p^{\dagger2}\big),\end{aligned}\end{equation}
where $\Delta_{p} = \omega_{p} - \omega_{d}/2$, $\Delta_{s} = \omega_{s} - \omega_{d}$, and $\widetilde{\Delta}_{m} = \omega_{m} - \omega_{d}$ are the detunings for the pump cavity, signal cavity, and magnon mode, respectively. We assume the condition $(\omega_{p},\omega_{d},\omega_{m})\ll\omega_{s}$, it is clear that the pump and magnon modes are dispersively coupled with the signal mode because of  the large frequency differences. To simplify the Hamiltonian further, we perfom a unitary transformation $U^{\prime}=\exp\left( i\Delta_{s} ta_{s}^{\dagger}a_{s}\right)$, and then obtain the following Hamiltonian
\begin{equation}\begin{aligned}H_I & =\Delta_{p}a_{p}^{\dagger}a_{p}+\widetilde{\Delta}_{m}m^{\dagger}m+J\big(a_{p}^2a_{s}^{\dagger}e^{i\Delta_{s}t}+a_{p}^{\dagger2}a_{s}e^{-i\Delta_{s}t}\big)\\  & \quad+g_{ms}\big(a_{s}m^{\dagger}e^{-i\Delta_{s}t}+a_{s}^{\dagger}me^{i\Delta_{s}t}\big)+F\big(a_{p}^2+a_{p}^{\dagger2}\big).
 \end{aligned}\end{equation}
\par Now we assume that $\Delta_{s}\gg\{g_{ms},J\}.$ From Eq. (4), it can be concluded that large detuned couplings occur among three bosonic modes, we could solve it based on the second-order perturbation theory \cite{James2007}. As a result, the signal cavity can be elimitated adiabatically
using the method of Ref. \cite{James2007} and the effective Hamiltonian $H_{eff}$ is as follows:
\begin{equation}\begin{aligned}
H_{eff}&=\Delta_pa_p^\dagger a_p+\Delta_mm^\dagger m+\chi a_p^{\dagger2}a_p^2+g\Big(a_p^{\dagger2}m+a_p^2m^\dagger\Big)\\&\quad+F\Big(a_p^2+a_p^{\dagger2}\Big),
\end{aligned}\end{equation}
with $\Delta_{m}=\widetilde{\Delta}_{m}-\frac{g_{ms}^{2}}{\Delta_{s}}$ being the effective detuning of the magnon mode, $g=-\frac{g_{ms}J}{\Delta_{s}}$ being the effective coupling strength between the pump and magnon mode, and $\chi=-\frac{J^{2}}{\Delta_{s}}$ being the Kerr nonlinearity. The dissipations of the cavity and magnon modes can be taken into account by an effective non-Hermitian Hamiltonian,
$H^{\prime}=H_{eff}-i\frac{\kappa}{2}a_p^{\dagger}a_p-i\frac{\gamma}{2}m^{\dagger}m,$
where $\kappa$ and $\gamma$ represent the decay rates of the two cavities.
We study the statistical properties of the magnon of this nonlinear quantum system, which can be described by a second-order correlation function, defined as \cite{ Scully1997}
\begin{equation}
g^{(2)}(0)=\frac{Tr(m^{\dagger2}m^2\rho)}{[Tr(m^\dagger m\rho)]^2}=\frac{\langle m^{\dagger2}m^2\rangle}{\langle m^\dagger m\rangle^2},
\end{equation}
where $m$ is the annihilation operator for magnon mode, and $\rho$ is the steady-state density matrix. Generally, $g^{(2)}(0)>1$ represents the magnon bunching effect, and $g^{(2)}(0)<1$ represents the magnon antibunching effect. The perfect magnon blockade effect is observed when $g^{(2)}(0)\rightarrow0$. In the following, we theoretically analyse and numerically calculate $g^{(2)}(0)$ of Eq. (7) by assuming realistic parameters.
\par Next, we search for optimal conditions using the probability amplitude method. We need to discuss the equal-time correlations on a truncated Fock state basis. In this system, we assume that the driving and coupling strengths are relatively weak. Here, we analytically solve the Schr\"odinger equation $i\partial|\Psi(t)\rangle/\partial t=H^{\prime}|\Psi(t)\rangle,$ where $\left|\Psi(t)\right\rangle $ is the time-dependent state of the present system. The general wave function of the system can be expanded in the few-photon and few-magneton subspace as \cite{Bamba2011,ZhangJS2019}:
\begin{equation}\begin{aligned}|\Psi(t)\rangle &= C_{00}|0,0\rangle+C_{01}|0,1\rangle+C_{20}|2,0\rangle+C_{40}|4,0\rangle\\&\quad+C_{21}|2,1\rangle+C_{02}|0,2\rangle,
\end{aligned}\end{equation}
where $|nm\rangle\equiv|n\rangle_{photon}\otimes|m\rangle_{\mathrm{magneton}},$ and $c_{n,m}$ is the probability amplitude. Under the weak driving condition, the coefficients should satisfy the conditions $C_{00}\gg\{C_{20},C_{01}\}\gg\{C_{40},C_{21},C_{02}\}$ and $C_{00}\approx1.$

Based on the Schr\"{o}dinger equation, the dynamical equations for the amplitudes can be written as \cite{Shen2014}
\begin{align}
        i\dot{C}_{00}  &  =\sqrt{2}FC_{20},\\
        i\dot{C}_{20}  &  =\left(  2\Delta_{p}+2\chi-i{\kappa}\right)  C_{01}%
        +\sqrt{2}FC_{00}+\sqrt{2}gC_{01}%
        \\&\quad +\sqrt{12}FC_{40},\nonumber \\
        i\dot{C}_{01}  &  =\left(\Delta_{m}-i\frac{\gamma}{2}\right)C_{01}+\sqrt{2}%
        gC_{20}+\sqrt{2}FC_{21},\\
        i\dot{C}_{40}  &  =\left(  4\Delta_{p}+12\chi-i2{\gamma}\right)  C_{40}%
        +\sqrt{12}gC_{21}+\sqrt{12}FC_{20},\\
        i\dot{C}_{21}  &  =\left[  2\Delta_{p}+\Delta_{m}-i\left(  \frac{\gamma}%
        {2}+\kappa\right)  \right]  C_{21}+\sqrt
        {12}gC_{40}+2gC_{02}
        \\&\quad +\sqrt{2}FC_{01},\nonumber \\
        i\dot{C}_{02}  &  =\left(  2\Delta_{m}-i{\gamma}\right)  C_{02}%
        +2gC_{21}.
\end{align}
\par When $\{\Delta_{p},\Delta_{m}\}\gg\{\kappa,\gamma\},$ the detunings are much larger than the dissipation rates of the cavity and magnon modes, we can easily obtain the steady-state solutions of the above equations, which yields
\begin{equation}C_{01}=-\frac{g{F}}{ g ^ 2 - ( \chi + \Delta _ p )  \Delta _ m },
\end{equation}
\begin{equation}C_{21}=-\frac{gF^2\Delta_mX}{\sqrt{2}Y(Z+g^2X)},
\end{equation}
\begin{equation}C_{02}=\frac{g^2F^2X}{\sqrt{2}Y(Z+g^2X)},
\end{equation}
with $X = 6\chi+2\Delta_p+3\Delta_m$, $Y = g^2-(\chi+\Delta_p)\Delta_m$, and $Z = -\Delta_m(3\chi+\Delta_p)(2\chi+2\Delta_p+\Delta_m).$
We obtain the steady-state second-order correlation function
\begin{equation}g_{mm}^{(2)}(0)\simeq\frac{2|C_{02}|^{2}}{|C_{01}|^{4}},\end{equation}
it is obvious that the condition $g_{mm}^{(2)}(0)={0}$ could be satisfied if $C_{02}={0}$. From Eq. (18), we see the second-order correlation function is
\begin{equation}g_{mm}^{(2)}(0)=\frac{X^2Y^2}{(Z+g^2X)^{2}}.
\end{equation}
\par There is MB in the case of $g_{mm}^{(2)}(0)={0}$.
From the analytical expression of Eq. (19), there are two optimal conditions. According to the method given in Ref. \cite{Hou2019}, we find the first optimal condition is $X = 6\chi+2\Delta_{p}+3\Delta_{m}=0.$ Here, we discover a surprising phenomenon that the coupling strength is absent in this optimal condition. It is known as UMB. To understand the physical mechanism of this phenomenon, we examine these two transition pathways for two magnon excitations again. As shown in Fig. 1(b), the two paths $\mathrm{(I)}|0,0\rangle\to|2,0\rangle\to|4,0\rangle\to|2,1\rangle\to|0,2\rangle $ and $\mathrm{(II)}|0,0\rangle\to|2,0\rangle\to|0,1\rangle\to|2,1\rangle\to|0,2\rangle$ lead to destructive interference. The destructive interference between two excitation channels suppresses the probability of the two-magnon state significantly and the UMB phenomenon appears.
\par The second condition is $Y = g^2-(\chi+\Delta_p)\Delta_m=0$. The interaction between the pump cavity and magnon results in a non-uniform energy spectrum in strongly nonlinear quantum systems. It is worth pointing out that the magnon-pump cavity coupling strength must be strong enough to observe the MB at the frequencies $g^{2}=(\chi+\Delta_{p})\Delta_{m}$.
\par By introducing the dissipation terms of the pump cavity and magnon, the dynamics of the present system can be described by the master equation \cite{Johansson2012,Johansson2013}
\begin{equation}\dot{\rho}=-i\big[H_{eff},\rho\big]+\frac{\kappa}{2}\mathcal{L}\big[a_{p}\big]\rho+\frac{\gamma}{2}\mathcal{L}[m]\rho,
\end{equation}
where $\rho$ is the density matrix of the system and $\mathcal{L}[\sigma]\rho=2\sigma\rho\sigma^{\dagger}-\sigma^{\dagger}\sigma\rho-\rho\sigma^{\dagger}\sigma $ is the Lindblad term accounting for losses to the environment.
\par As shown in Fig. 2, the numerical simulations agree with the analytical solutions for the relevant parameters $\Delta_{p}$ in different cases, confirming the above calculations.
\begin{figure}[tbp]
\centering {\scalebox{0.29}[0.29]{\includegraphics{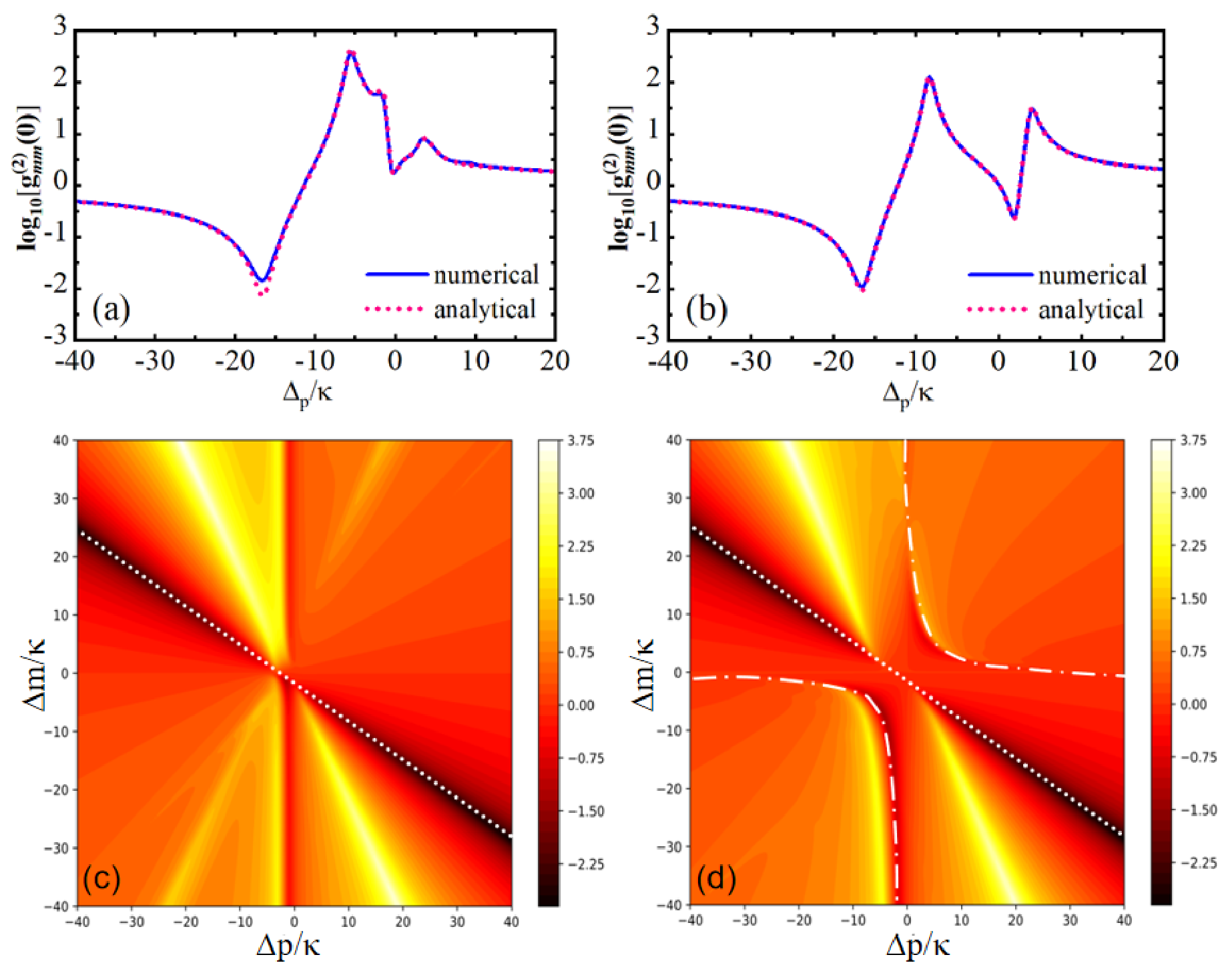}}}
\caption{The analytical solutions and the numerical results of $\log_{10}g^{(2)}_{mm}(0)$ are plotted as functions of the detuning $\Delta_p/\kappa$; $\chi=0.5\kappa$ and $g=0.5\kappa$ in (a); $\chi=0.5\kappa$ and $ g=5\kappa$ in (b). Equal-time second-order correlation function $\log_{10}g^{(2)}_{mm}(0)$ versus the photon detuning $\Delta_p$ and magnon detuning $\Delta_m$, the coupling strength is chosen as (c) $g=0.5\kappa$ and (d) $g=5\kappa$. The white dotted lines represent the the interference-induced UMB condition ($6\chi+2\Delta_p+3\Delta_m=0$) and the white dash-dotted curves denote the condition of MB ($g^{2}=(\chi+\Delta_{p})\Delta_{m}$). The other parameters are $\kappa/2\pi=2.0\mathrm{~MHz}, \gamma/2\pi=1.0\mathrm{~MHz}$, $\Delta_{m}=10\kappa$, and $\mathrm{F}=0.05\kappa$.}\label{fig2}
\end{figure}
In Fig. 2(a), for weak coupling strength $g=0.5\kappa$, we find the minimum value of $g^{(2)}_{mm}(0)$ reaches below $10^{-2}$, resulting in a strong UMB effect. In Fig. 2(b), for strong coupling strength $g=5\kappa$, it is clear to see two troughs in the second-order correlation function, corresponding to the frequencies $6\chi+2\Delta_{p}+3\Delta_{m}=0$ and $g^{2}=(\chi+\Delta_{p})\Delta_{m}$.
The left is the result of an unconventional solution, which is resulted from the quantum destructive interference, and the right one is attributed to the anharmonic energy splitting, which is sensitive to the coupling strength. We see the value of $g^{(2)}_{mm}(0)$ at $6\chi+2\Delta_{p}+3\Delta_{m}=0$ is much smaller than that at $g^{2}=(\chi+\Delta_{p})\Delta_{m}$. Therefore, there is UMB even in the weak coupling regime and the scheme proposed here is feasible in experiments.
In Fig. 2(c) and Fig. 2(d), we plot the second-order correlation function $g^{(2)}_{mm}(0)$ as functions of the detunings $\Delta_{p}$ and $\Delta_{m}$. We choose the coupling strengths $g=0.5\kappa$ [Fig. 2(c)] and $g=5\kappa$ [Fig. 2(d)], respectively. It is evident that the second-order correlation function is consistently below unity at $6\chi+2\Delta_p+3\Delta_m=0$ in both weak and strong coupling regimes. Nevertheless, the correlation function at $g^{2}=(\chi+\Delta_{p})\Delta_{m}$ drops lower than unity only when the system transitions into the strong coupling regime, aligning closely with the analytical results.
\begin{figure}[tbp]
\centering {\scalebox{0.24}[0.26]{\includegraphics{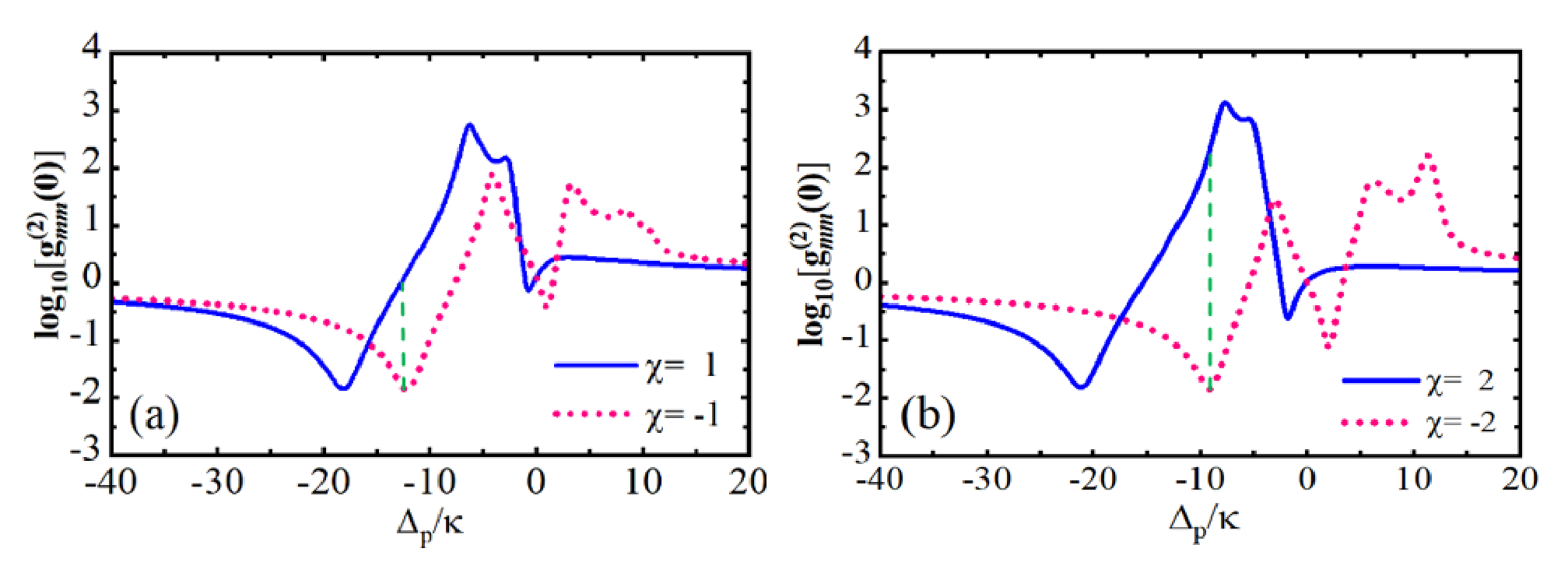}}}
\caption{The red dotted lines are the second-order correlation functions for positive Kerr effect coefficient. The blue solid lines are the second-order correlation functions for negative Kerr effect coefficient, $\log_{10}g^{(2)}_{mm}(0)$ are plotted as functions of the detuning $\Delta_p/\kappa$; $\chi=1\kappa$ (blue solid line) and $\chi=-1\kappa$ (red dotted line) in (a); $\chi=2\kappa$ (blue solid line) and $\chi=-2\kappa$ (red dotted line) in (b). The other parameters are $\kappa/2\pi=2.0\mathrm{~MHz}$, $\gamma/2\pi=1.0\mathrm{~MHz}$, $\Delta_m=10\kappa$, $\mathrm{F}=0.05\kappa$, and $g=1\kappa$.}\label{fig3}
\end{figure}

In Fig. 3, we numerically simulate the second-order correlation function $\log_{10}g^{(2)}_{mm}(0)$ as functions of the detunings $\Delta_{p}$ and $\Delta_{m}$ for different values of $\chi$. From Eq. (6), we find $\chi=-\frac{J^{2}}{\Delta_{s}}$ can be positive or negative since one can adjust the detunning $\Delta_{s}$ conveniently. We know that nonreciprocal devices allow the propagation of signals from one side but not from the other. Very recently, NPB have been realized in a spinning optomechanical system \cite{Li2019}, and a driven dissipative cavity with parametric amplification \cite{Shen2020}. This nonreciprocity results from the Sagnac-Fizeau shift in the spinning system, leading to strong UPB only by driving the device from one side but not from the other side. In addition, a nonreciprocal magnonic device that can be experimentally realized at the single-magnon level was proposed in Ref. \cite{Wang2022}. It was pointed out that the interference between the coherent and dissipative qubit-magnon couplings can yield the NMB \cite{Wang2022}. Different from the previous work, we achieve NUMB here by changing the positivity and negativity of the Kerr nonlinearity parameter. In Fig. 3(a), there is a green dashed line. A deep dip at the frequency $6\chi+2\Delta_{p}+3\Delta_{m}=0$ can be observed in the second-order correlation function. When $\Delta_p/\kappa = -12$, the value of $g^{(2)}_{mm}(0)$ is close to $10^{-2}$ due to the destructive interference, resulting in a strong UMB effect. Moreover, the top of the green dashed line corresponds to the positive value of $\chi$, it is evident that there is no blockade effect. It is apparent that this phenomenon is asymmetric, which proves the NUMB. The nonreciprocal effect is enhanced significantly as the absolute values of the Kerr parameters increase. The optimal results of $g^{(2)}_{mm}(0)=0$ are presented with $\Delta_{p}/\kappa$ in Fig. 3, satisfying optimal analytical and numerical conditions. These results demonstrate that the analytical calculations for the nonreciprocal magnon antibunching are in excellent agreement with those obtained by the numerical simulations.
\par In summary, we have presented to combine the nonlinear photon-photon coupling and magnetic dipole interaction to achieve NUMB in a hybrid nonlinear system formed by a pump cavity, a signal cavity and a YIG sphere. The optimal parameter conditions to achieve perfect UMB are obtained by calculating the analytical expression of the second-order correlation function. The destructive interference between different transition paths causes the UMB effect. The numerical simulation results are highly consistent with the analytical solution. In addition, we have shown that the nonreciprocity is due to the adjustable Kerr parameters, which can be positive or negative by adjusting the detuning. Our work opens up a new way to achieve NMB and provides the possibility of generating single magnon source in quantum information processing.
\section*{Funding.}
National Natural Science Foundation of China (11905064, 12165007, 12175199); Natural Science Foundation of Jiangxi Province (20232ACB201013).

\section*{Disclosures.} The authors declare no conflicts of interest.


\begin{thebibliography}{99}

\bibitem{Imamoglu1997} A. Imamo\=glu, H. Schmidt, G. Woods, and M. Deutsch, "Strongly interacting photons in a nonlinear cavity," Phys. Rev. Lett. \textbf{79}, 1467(1997).

\bibitem{Miranowicz2013} A. Miranowicz, M. Paprzycka, Y. X. Liu, J. Bajer, and F. Nori, "Two-photon and three-photon blockades in driven nonlinear systems," Phys. Rev. A \textbf{87}, 023809 (2013).


\bibitem{Birnbaum2005} K. M. Birnbaum, A. Boca, R. Miller, A. D. Boozer, T. E. Northup, and H. J. Kimble, "Photon blockade in an optical cavity with one trapped atom," Nature \textbf{436}, 87 (2005).

\bibitem{Hoffman2011} A. J. Hoffman, S. J. Srinivasan, S. Schmidt, L. Spietz, J. Aumentado, H. E. T\"ureci, and A. A. Houck, "Dispersive photon blockade in a superconducting circuit," Phys. Rev. Lett. \textbf{107}, 053602 (2011).


\bibitem{Faraon2008} A. Faraon, I. Fushman, D. Englund, N. Stoltz, P. Petroff, and J. Vu\v ckovi\'c, "Coherent generation of non-classical light on a chip via photon-induced tunnelling and blockade," Nat. Phys. \textbf{4}, 859 (2008).


\bibitem{Shen2015} H. Z. Shen, Y. H. Zhou, and X. X. Yi, "Tunable photon blockade in coupled semiconductor cavities," Phys. Rev. Lett. \textbf{91}, 063808 (2015).

\bibitem{Liew2010}T. C. H. Liew and V. Savona, "Single photons from coupled quantum modes," Phys. Rev. Lett. \textbf{104}, 183601 (2010).

\bibitem{Flayac2013}H. Flayac and V. Savona, "Input-output theory of the unconventional photon blockade," Phys. Rev. A \textbf{88}, 033836 (2013).

\bibitem{Flayac2017}H. Flayac and V. Savona, "The Unconventional Photon Blockade," Phys. Rev. A \textbf{96}, 053810 (2017).



\bibitem{Bamba2011} M. Bamba, A. Imamo\=glu, I. Carusotto, and C. Ciuti, "Origin of strong photon antibunching in weakly nonlinear photonic molecules," Phys. Rev. A \textbf{83}, 021802(R) (2011).

\bibitem{Kyriienko2014} O. Kyriienko, I. A. Shelykh, and T. C. H. Liew, "Tunable single-photon emission from dipolaritons," Phys. Rev. A \textbf{90}, 033807 (2014).

\bibitem{Zhou2015} Y. H. Zhou, H. Z. Shen, and X. X. Yi, "Unconventional photon blockade with second-order nonlinearity," Phys. Rev. A \textbf{92}, 023838 (2015).


\bibitem{Xu2014} X. W. Xu and Y. Li, "Tunable photon statistics in weakly nonlinear photonic molecules," Phys. Rev. A \textbf{90}, 043822 (2014).


\bibitem{Sarma2018} B. Sarma and A. K. Sarma, "Unconventional photon blockade in three-mode optomechanics," Phys. Rev. A \textbf{98}, 013826 (2018).


\bibitem{Wang2023} F. Wang and C. Gou, "Magnon-induced absorption via quantum interference," Opt. Lett. \textbf{48}, 1164 (2023).

\bibitem{Bourhill2016} J. Bourhill, N. Kostylev, M. Goryachev, D. L. Creedon, and M. E. Tobar, "Ultrahigh cooperativity interactions between magnons and resonant photons in a YIG sphere," Phys. Rev. B \textbf{93}, 144420 (2016).

\bibitem{Liu2019} Z. X. Liu, H. Xiong, and Y. Wu, "Magnon blockade in a hybrid ferromagnet-superconductor quantum system," Phys. Rev. B \textbf{100}, 134421 (2019).


\bibitem{Wu2021} K. Wu, W. X. Zhong, G. L. Cheng, and A. X. Chen, "Phase-controlled multimagnon blockade and magnon-induced tunneling in a hybrid superconducting system," Phys. Rev. A \textbf{103}, 052411 (2021).
\bibitem{Li2018} J. Li, S. Y. Zhu, and G. S. Agarwal, "Magnon-photon-phonon entanglement in cavity magnomechanics," Phys. Rev. Lett. \textbf{121}, 203601 (2018).
\bibitem{Zheng2023} Q. J. Zheng, W. X. Zhong, G. L. Cheng, and A. X. Chen, "Nonreciprocal tripartite entanglement based on magnon Kerr effect in a spinning microwave resonator," Opt. Commun. \textbf{546}, 129796 (2023).

\bibitem{Chen2023} J. J. Chen, X. G. Fan, W. Xiong, D. Wang, and L. Ye, "Nonreciprocal entanglement in cavity-magnon optomechanics," Phys. Rev. B \textbf{108}, 024105 (2023).

\bibitem{Xie2020} J. K. Xie, S. L. Ma, and F. L. Li, "Quantum-interference-enhanced magnon blockade in an yttrium-iron-garnet sphere coupled to superconducting circuits," Phys. Rev. A \textbf{101}, 042331 (2020).

\bibitem{Fan2023} Y. Q. Fan, J. H. Li, and Y. Wu, "Nonclassical magnon pair generation and Cauchy-Schwarz inequality violation," Phys. Rev. A \textbf{108}, 053715 (2023).

\bibitem{WangF2022} F. Wang, C. D. Gou, J. Xu, and C. Gong, "Hybrid magnon-atom entanglement and magnon blockade via quantum interference," Phys. Rev. A \textbf{106}, 013705 (2022).

\bibitem{Yan2024} Y. T. Yan, C. S. Zhao, D. W. Wang, J. Y. Yang, and L. Zhou, "Simultaneous blockade of two remote magnons induced by an atom," Phys. Rev. A \textbf{109}, 023710 (2024).


\bibitem{Bino2018} L. D. Bino, J. M. Silver, M. T. M. Woodley, S. L. Stebbings, X. Zhao, and P. DelHaye, "Microresonator isolators and circulators based on the intrinsic nonreciprocity of the Kerr effect," Optica. \textbf{5}, 279 (2018).

\bibitem{Xie2022} H. Xie, L. W. He, X. Shang, G. W. Lin, and X. M. Lin, "Nonreciprocal photon blockade in cavity optomagnonics," Phys. Rev. A \textbf{106}, 053707 (2022).

\bibitem{Huang2024} K. W. Huang, X. Wang, Q. Y. Qiu, and H. Xiong, "Nonreciprocal magnon blockade via the Barnett effect," Opt. Lett. \textbf{49}, 758 (2024).

\bibitem{Fang2023} Y. J. Fang , W. X. Zhong, G. L. Cheng, and A. X. Chen, "Magnon-photon cross-correlations via optical nonlinearity in cavity magnonical system," Opt. Express \textbf{31}, 27381 (2023).
\bibitem{James2007} D. F. James and J. Jerke, "Effective hamiltonian theory and its applications in quantum information," Can. J. Phys. \textbf{85}, 625 (2007).
\bibitem{Scully1997} M. O. Scully and M. S. Zubairy, \emph{Quantum Optics} (Cambridge University Press, 1997).
\bibitem{ZhangJS2019} J. S. Zhang, M. C. Li, and A. X. Chen, "Enhancing quadratic optomechanical coupling via a nonlinear medium and lasers," Phys. Rev. A \textbf{99}, 013843 (2019).

\bibitem{Shen2014} H. Z. Shen, Y. H. Zhou, and X. X. Yi, "Quantum optical diode with semiconductor microcavities," Phys. Rev. A \textbf{90}, 023849 (2014).

\bibitem{Hou2019} K. Hou, C. J. Zhu, Y. P. Yang, and G. S. Agarwal, "Interfering pathways for photon blockade in cavity QED with one and two qubits," Phys. Rev. A \textbf{100}, 063817 (2019).

\bibitem{Johansson2012} J. R. Johansson, P. D. Nation, and F. Nori, "QuTiP: An open-source Python framework for the dynamics of open quantum systems," Comp. Phys. Com. \textbf{183}, 1760 (2012).
\bibitem{Johansson2013} J. R. Johansson, P. D. Nation, and F. Nori, "QuTiP 2: A Python framework for the dynamics of open quantum systems," Comp. Phys. Com. \textbf{184}, 1234 (2013).

\bibitem{Li2019} B. Li, R. Huang, X. Xu, A. Miranowicz, and H. Jing, "Nonreciprocal unconventional photon blockade in a spinning optomechanical system," Photon. Res. \textbf{7}, 630 (2019).

\bibitem{Shen2020} H. Z. Shen, Q. Wang, J. Wang, and X. X. Yi, "Nonreciprocal unconventional photon blockade in a driven dissipative cavity with parametric amplification," Phys. Rev. A \textbf{101}, 013826 (2020).

\bibitem{Wang2022} Y. Wang, W. Xiong, Z. Xu, G. Q. Zhang, and J. Q. You, "Dissipation-induced nonreciprocal magnon blockade in a magnon-based hybrid system," Sci. China Phys. Mech. Astron. \textbf{65}, 260314 (2022).

\end{thebibliography}

\newpage
\end{document}